\documentclass[doublecol]{epl2} %for 2 columns style without line numbers
\usepackage{enumitem}
\usepackage{amsmath}
\usepackage{amssymb}

\renewcommand{\Re}{\operatorname{Re}}

\title{Tensional buckling around an elliptic hole in a stretched sheet}

\author{I. Andrade-Silva \and M. Adda-Bedia}

\institute{                    
Universit\'e de Lyon, Ecole Normale Sup\'erieure de Lyon, Universit\'e Claude Bernard,
CNRS, Laboratoire de Physique, F-69342 Lyon, France
}


\abstract{
It is well known that an annular sheet could wrinkle as a result of axisymmetric tensile loads applied at the edges. In this system, regions under compression appear due to Poisson effect in the azimuthal direction yielding an incompatible excess in length, so that the membrane has to buckle out of the plane. Then, radial wrinkles emerge following the direction of the tensile principal stress. This so called Lam\'e configuration has been widely used as theoretical and experimental paradigms for wrinkling instabilities. In this work, we explore the consequences of changing the geometry of this model configuration by considering an elliptic hole in an infinite stretched membrane. Using the Kolosoff-Inglis solution, we analyse the stress field around the hole and identify three possible regions: taut, unidirectionally tensioned and slack regions. According to these definitions, we classify in a phase diagram the different stress states of the sheet as function of the hole eccentricity and of the applied tensions. Finally, we quantify how the tension lines vary with the geometry of the hole and discuss possible outcomes on the emerging buckled patterns.
}

\begin{document}

\maketitle

Wrinkle and fold formation in elastic thin plates has aroused a great interest due to its occurrence in a wide range of natural and manmade systems~\cite{borges1960scar,burton1999keratocytes,cerda2003geometry,lacour2003stretchable,geminard2004wrinkle,cerda2005mechanics,davidovitch2011prototypical,davidovitch2012nonperturbative,deng2016wrinkled}. In biophysics, particular attention has been paid in understanding traction forces involved in cell motility on elastic substrates using the emerging wrinkling patterns~\cite{burton1999keratocytes}. For medical purposes, a better understanding of wrinkle formation would be helpful for the treatment of post-surgery scars~\cite{cerda2005mechanics,borges1960scar}. In condensed matter physics, pre-wrinkled conductive films have been proposed as stretchable electric contacts~\cite{lacour2003stretchable} and wrinkles in graphene sheets are believed to modify its electronic properties~\cite{deng2016wrinkled}. Finally, examples of wrinkled patterns are abundant in daily life: from clothes or curtains to human skin or fruit peels~\cite{cerda2005mechanics}. Wrinkles in thin elastic films emerge as a buckling instability due to in-plane compressive stresses. However, wrinkles can also be created when a sheet is subjected to a large enough longitudinal stretching yielding a transverse compressive stress due to the Poisson effect. 
In this case, the film buckles to relax the in-plane strain incompatibility producing out-of-plane wavy deformations. 

When wrinkling is induced by in-plane tension in membranes of zero flexural rigidity, the direction of wrinkles is described by tension field theory~\cite{reissner,mansfield2005bending} which assumes that the wrinkles occur along tension rays parallel to the direction of the largest principal stress while the smallest principal stress collapses to zero. These assumptions are supported by the fact that a plate with zero flexural stiffness cannot sustain compressive stresses, the so-called membrane limit. Tension field theory has been applied to predict wrinkle directions in some specific geometries~\cite{danielson1975tension}, however this theory, as it was formulated originally, is not able to predict the fine features of wrinkle patterns such as their extension, wavelength and amplitude.

When the flexural stiffness is small but finite, one can identify basically two regimes: a near-threshold (NT) regime which is just beyond the onset of buckling instability~\cite{geminard2004wrinkle} and a far-from-threshold (FFT) regime in which the wrinkles are well developed~\cite{davidovitch2011prototypical}. The NT regime can be characterised by a perturbative analysis of the F\"oppl-von Karman (FvK) equations over the flat state.  Then, one can derive scaling laws for the wavelength and extension of the wrinkles. In the FFT regime, wrinkling induces a collapse of the stress component in the direction of compression~\cite{davidovitch2012nonperturbative}. Upon this assumption, the extension of the wrinkles and their wavelength are obtained by minimising an energy functional. Thus, FFT analysis is a tension field theory for membranes with finite flexural rigidity~\cite{davidovitch2012nonperturbative}.

Both NT and FFT analysis can be performed when the direction of the subsequent wrinkles is known a priori, which is the case of simple geometries such as rectangular~\cite{cerda2002thin} or annular sheets (also known as Lam\'e configuration)~\cite{vella2010capillary}. However, little is known when the lines of tension in the pre-buckled state of a sheet are not straight. Slight changes on the boundary conditions, such as nonuniform tensions or asymmetric geometries would lead to curved tension lines. The characterisation of the wrinkling instability for general geometrical and loading conditions remains an open problem~\cite{aharoni2017smectic}. 

The present work explores the role of geometry on the tensional wrinkling instabilities. As a natural extension of the Lam\'e configuration, we study the elastic problem of an infinite elastic sheet perforated by an elliptic hole and subjected to a uniform differential tension between its outer boundary and inner edge. We obtain the stress field for this configuration and compute both its principal components and principal directions as function of the eccentricity of the hole and the ratio between the inward and outward tensions. We notice that, depending on these control parameters, regions around the membrane can be either in a taut, slack or unidirectionally tensional state. This yields a rich phase diagram of different global stress states depending on the topology of the compressed regions and on the presence (or absence) of slack regions. Finally, we compute the lines of tensions and show that as one moves away from the circular hole case towards the crack limit, lines of tensions bent considerably, making it difficult to predict wrinkling patterns beyond the buckling instability.

\begin{figure}[t]
\begin{center}
\includegraphics[width=0.6\columnwidth]{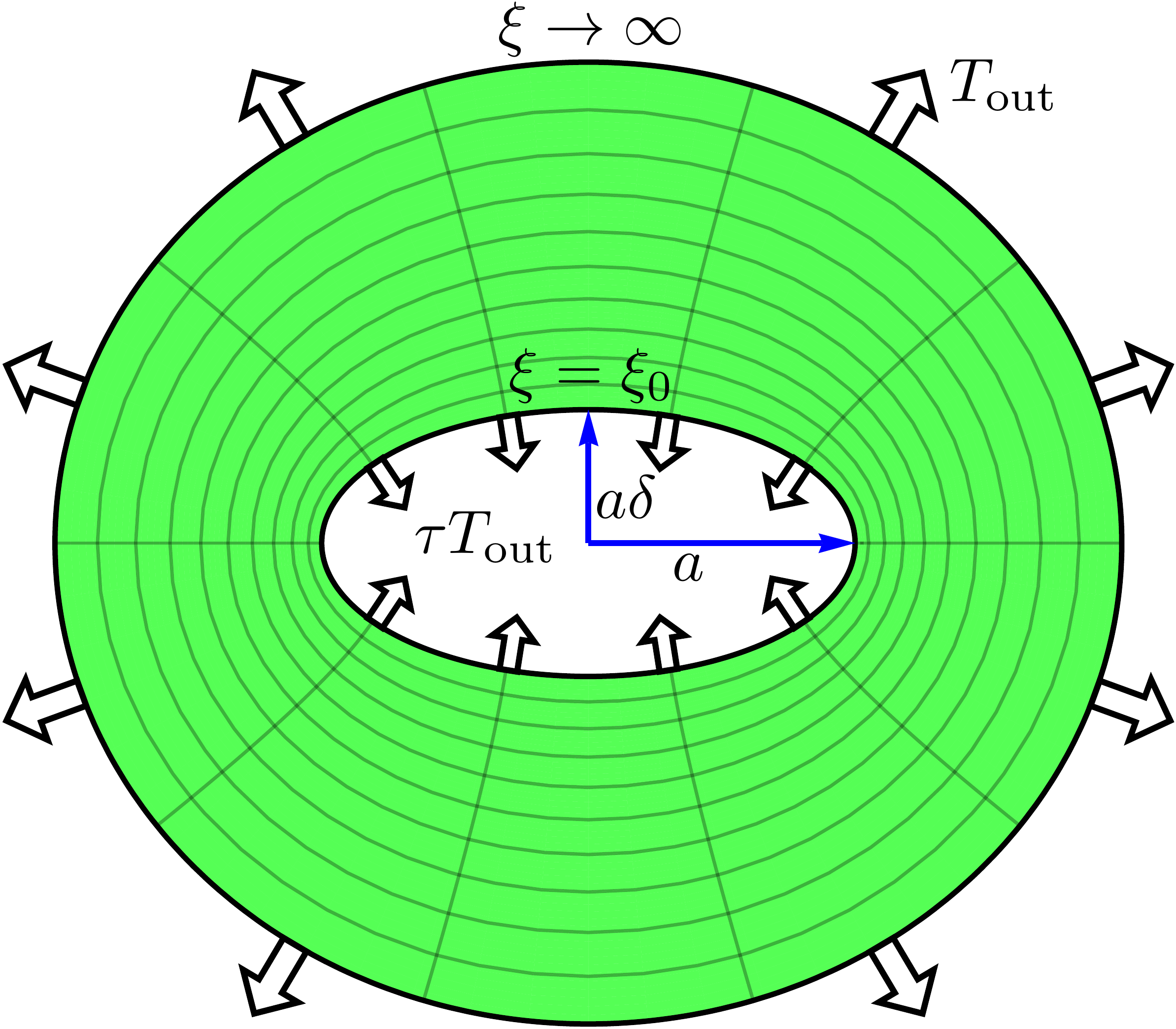}
\end{center}
\caption{Schematics of the modified Lam\'e problem. Lines on the sheet illustrate the elliptic coordinate system $(\xi,\eta)$.}
\label{Fig:scheme}
\end{figure}

\section{Lam\'e problem for an elliptic hole}

Fig.~\ref{Fig:scheme} shows an infinite sheet with an elliptic hole of semi axes $a$ and $a \delta$, with $\delta<1$, that is stretched through uniform tensions applied at its boundaries. For this configuration, it is suitable to use the elliptic coordinates $\xi>0$ and $\eta \, \in [0,2\pi ]$ given by
\begin{equation}
\left(x,y \right) = \left(c \, \cosh \, \xi \cos \, \eta, c \, \sinh \, \xi \sin \, \eta  \right),
\label{eq:transf}
 \end{equation}
where $\left(x,y \right)$ are the Cartesian coordinates whose origin is at the centre of the hole, $c=a\sqrt{1-\delta^2}$ and the boundary of the inner hole corresponds to the ellipse $\xi=\xi_0 = \tanh^{-1}\delta$. The sheet is subjected to tensions applied at the boundaries: a uniform all-around tension $T_{\text{out}}$ at an infinite distance from the hole and an inward tension $\tau T_{\text{out}}$, with $\tau >0$, at the inner edge. Thus, the boundary conditions can be written as follows
\begin{eqnarray}
&& \text{at } \xi \rightarrow \infty\text{ : } \sigma_{\xi \xi} = \sigma_{\eta \eta} = T_{\text{out}} , \label{Boundary condition 1}\\
&& \text{at } \xi=\xi_0\text{ : } \sigma_{\xi \xi}= \tau T_{\text{out}}, \text{ and } \sigma_{\xi \eta} = 0  . \label{Boundary condition 2}
\end{eqnarray}

The corresponding elastic problem can be solved using the superposition principle. First, we consider the Kolosoff-Inglis solution~\cite{kolosoff1913solution} of a sheet subjected to a uniform pressure $-\Delta T$, with $\Delta T = (\tau-1)T_{\text{out}}$, at infinity while the boundary of the hole is traction-free. The associated stress field of this classical problem is reported in~\cite{timoshenko412theory}. Then to satisfy the boundary conditions~(\ref{Boundary condition 1},\ref{Boundary condition 2}), we take advantage of the linearity of the system and add to the latter solution a homogeneous stressed state of the sheet given by $\sigma_{\xi \xi} =\sigma_{\eta \eta} = \tau T_{\text{out}}$ and $\sigma_{\xi \eta}=0$. One can show that the components of the resulting stress tensor are given by~\cite{timoshenko412theory}
\begin{subequations}
\begin{eqnarray}
\sigma_{\xi \xi} + \sigma_{\eta \eta} &=& 2\tau T_{\text{out}} - 2 \Delta T \Re\left[\coth{\zeta}\right], \\
\sigma_{\eta \eta} -\sigma_{\xi \xi} + 2 i \sigma_{\xi \eta} &=& \Delta T\, \frac{\cosh{\bar{\zeta}}-\cosh{2\xi_0}\cosh{\zeta}}{\sinh{\bar{\zeta}} \sinh^2{\zeta}},
\label{Shear stress xi eta} 
\end{eqnarray}
\label{stresses}
\end{subequations}
where $\zeta=\xi+i\eta$.
In the following, we study the behaviour of the stress field as given by Eqs.~(\ref{stresses}) and draw conclusions on the state of the membrane. To identify regions under compression, one should determine the principal stresses and their associated principal directions. After some algebraic manipulations, one computes the largest and smallest principal stresses, $\sigma_{1}$ and $\sigma_{2}$, as well as the angle $\beta$ between the $x$-axis and the direction of the principal stresses~\cite{adda1996morphological}. The results are  given by Eqs.~(\ref{principal-stresses}).
\begin{widetext}
\begin{subequations}
\begin{eqnarray}
\sigma_{1,2} &=&\tau T_{\text{out}}-\frac{\Delta T\sinh 2\xi}{ \cosh 2\xi -\cos 2 \eta}  \pm |\Delta T| \frac{\sqrt{ (\cosh 2\xi - \cosh 2\xi_0 )^{2 } \sin^{2} 2 \eta +( \cosh 2 \xi_{0} - \cos 2\eta)^{2}\sinh^{2} 2 \xi } }{( \cosh 2\xi -\cos 2 \eta )^{2}}, \label{Principal stress}\\
\tan \, 2\beta &=& \frac{(\sigma_{\xi \xi} - \sigma_{\eta \eta})\sinh 2\xi \sin 2\eta   + 4\sigma_{\xi \eta} (\sinh^2\xi-\sin^2\eta)}{2(\sigma_{\xi \xi} - \sigma_{\eta \eta})(\sinh^2\xi-\sin^2\eta) -2 \sigma_{\xi \eta} \sinh 2\xi \sin 2\eta}, \label{tan 2 beta} 
\end{eqnarray}
\label{principal-stresses}
\end{subequations}
\end{widetext}
 Notice that $\beta$ does not depend explicitly on the applied loadings and thus, is a function of the geometrical parameter $\delta$ only.

\begin{figure}[t]
\begin{center}
\includegraphics[width=\columnwidth]{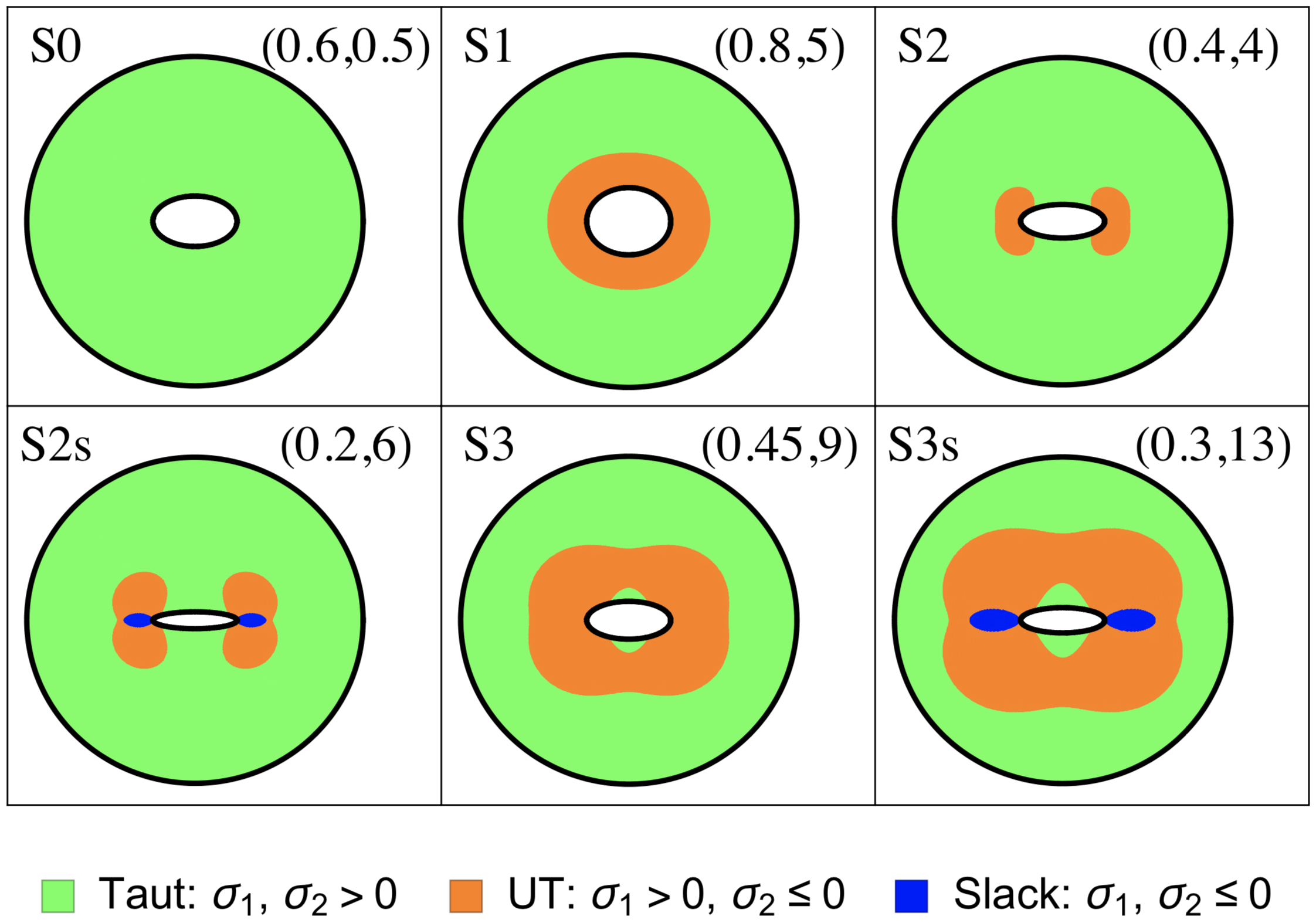}
\end{center}
\caption{\label{Fig: FiguresSheet}  Representation of the six possible states of the stretched membrane, with taut, UT and slack regions. The corresponding values of the control parameters $(\delta,\tau)$ are shown in each panel.}
\end{figure}

\begin{figure}[t]
\begin{center}
\includegraphics[width=\columnwidth]{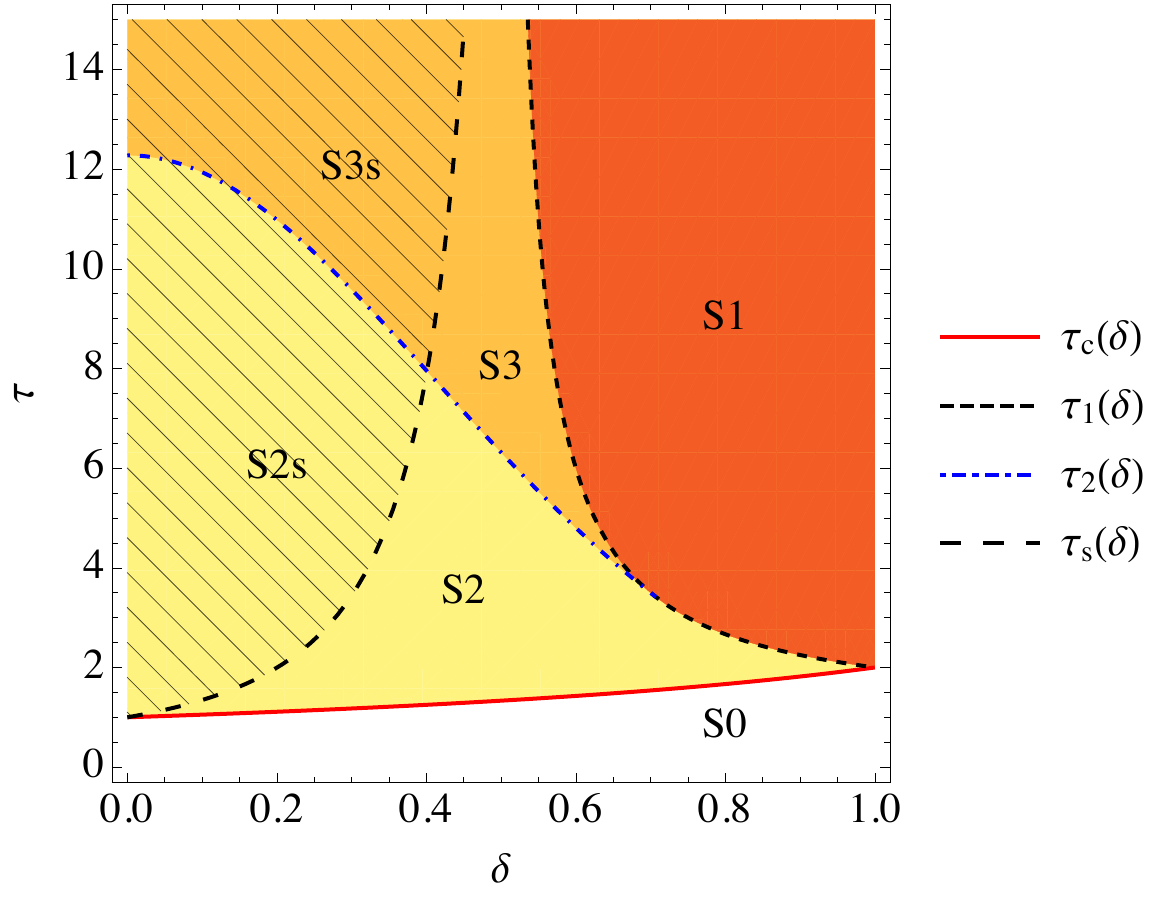}
\end{center}
\caption{\label{Fig: Diagram} Diagram showing the different states shown in Fig.~\ref{Fig: FiguresSheet} in the parameter space $(\delta,\tau)$. The critical curves $\tau_c(\delta)$, $\tau_{1}(\delta)$, $\tau_{2}(\delta)$ and $\tau_{s}(\delta)$ delimiting the different regions are defined in the text. }
\end{figure}

Perturbative analysis of tensional wrinkling usually focuses on the pre-buckled planar state of the membrane to define the regions of the membrane where a single component of the stress tensor is compressive. In the present case, both principal stresses can be either negative or positive in different regions. Hence, when $\sigma_{1}$ reaches negative values, the membrane becomes slack. In order to analyse in detail the stress state on the membrane, we classify regions on the sheet according to the signs of their principal stresses :
\begin{itemize}[label=$\bullet$,leftmargin=*]
\item Taut region: $\sigma_{1}>0,\,\sigma_{2} > 0$.
\item Unidirectional stretched (UT) region:
$\sigma_{1} > 0,\,\sigma_{2} \le 0$.
\item Slack region: $\sigma_{1}\le0,\, \sigma_{2}\le 0$.
\end{itemize}
Fig.~\ref{Fig: FiguresSheet} shows all the possible states of the membrane for different values of the parameters ($\delta,\tau$). Far enough from the hole, the sheet is always in a taut state. However, depending on the values of the control parameters, regions of compressive stress(es) will appear around the hole. UT regions can either concentrate at the ends of the major axis of the elliptic hole or surround it entirely (as in the Lam\'e case). Slack regions can also appear as small elliptical spots around the tips of the ellipse.  A relevant question for a wrinkling stability analysis is whether the regions under compression that surround the hole are connected or disconnected. According to Fig.~\ref{Fig: FiguresSheet}, one can identify six possible global states that can characterised by the values of the principal stresses at some specific locations:
\begin{enumerate}[label=$\bullet$,leftmargin=*]
\item S0: the entire sheet is taut: 
$\sigma_{2} (\xi_{0},\eta)>0,\, \forall \, \eta \in [0,2\pi]$. 
\item S1: a single UT region surrounding the hole such that each point at the edge is under compression: $\sigma_{2}(\xi_{0},\eta) < 0,\, \forall\, \eta \in [0,2\pi]$ and $\sigma_{1}>0$ everywhere. 
\item S2: there are two disconnected UT regions at both ends of the major axis of the hole, without slack regions: $\sigma_{2}(\xi_{0},0) <0$ and $\sigma_{2}(\xi,\pi/2) >0 ,\, \forall \xi \ge \xi_{0}$.
\item S2s: similar to S2 but with the presence of slack regions around the tips of the ellipse.
\item S3: a single connected UT region around the ellipse without slack regions and with taut regions touching both ends of the minor axis of the hole: $\sigma_2 (\xi,\pi/2)\leq0$ in a finite interval $\xi_0<\xi_{\text{min}}\leq\xi\leq\xi_{\text{max}}$ and $\sigma_2 (\xi,\pi/2)>0$ elsewhere.
\item S3s: similar to S3 but with the presence of slack regions around the tips of the ellipse.
\end{enumerate}
Fig.~\ref{Fig: Diagram} shows the phase diagram of these different states in the parameter space $(\delta,\tau)$. The critical separation curves can be computed analytically as follows.

Due to the symmetries of the states shown in Fig.~\ref{Fig: FiguresSheet}, the phase diagram can be retrieved by examining the principal stresses along $\eta=0$ and $\eta=\pi/2$ only. Using Eqs.~(\ref{principal-stresses}), one has for $\tau>1$
\begin{eqnarray}
\sigma_{1,2}(\xi,0) &=&\tau T_{\text{out}}-\Delta T\coth\xi  \left[1\mp  \frac{ \sinh^2 \xi_{0} }{\sinh^2\xi}\right], \label{Principal stress_x}\\
\sigma_{1,2} (\xi,\pi/2)&=&\tau T_{\text{out}}-\Delta T\tanh \xi  \left[1\mp  \frac{\cosh^2 \xi_{0} }{\cosh^2\xi}\right]. \label{Principal stress_y}
\end{eqnarray}
Recall that $\delta=\tanh\xi_0$ and $\Delta T = (\tau-1)T_{\text{out}}$. At the boundary of the hole, one has $\sigma_{1}(\xi_0,0) =\sigma_{1}(\xi_0,\pi/2) =\tau T_{\text{out}}$, while $\sigma_{2}(\xi_0,0) \leq\sigma_{2}(\xi_0,\pi/2) $. Therefore, the threshold of apparition of compressive stresses around the hole is given by $\sigma_{2}(\xi_0,0) <0$. Using Eq.~(\ref{Principal stress_x}), this condition yields  
\begin{equation}
\tau>\tau_c(\delta) = \frac{2}{2-\delta}. \label{Curve tau_c}
\end{equation}
The region S0 is defined by the condition  $\tau<\tau_{c}(\delta)$ for which no compressive stresses occur. Notice that one recovers the threshold $\tau_c(1)=2$ corresponding to the Lam\'e problem~\cite{davidovitch2011prototypical} and that $\tau_c(0)=1$ for a crack-like interface. 
Region S1 is characterised by the existence of a compressive region all around the hole. Thus the curve $\tau_{1} (\delta)$ bounding S1 is found by imposing the condition $\sigma_{2}(\xi_0,\pi/2) <0$. Using Eq.~(\ref{Principal stress_y}), we obtain
\begin{equation}
\tau>\tau_{1} (\delta)= \frac{2\delta}{2\delta-1}. \label{Second inequality}
\end{equation} 
As expected from the Lam\'e case, one has $\tau_{1}(1)=\tau_{c}(1)=2$. Moreover, Eq.~(\ref{Second inequality}) shows that $\tau_{1}$ diverges as $\delta \rightarrow 1/2$.
 
To discriminate between the regions S2~[S2s] and S3~[S3s], one should explore the behaviour of $\sigma_2$ along the $y$-axis. In these regions, one has $\sigma_2 (\xi_0,\pi/2)>0$ and $\sigma_2 (\infty,\pi/2)>0$ but $\sigma_2 (\xi,\pi/2)$ may not be a monotonic function of $\xi\geq\xi_0$. The critical curve $\tau_{2}(\delta)$ separates a region S2~[S2s] in which $\sigma_2 (\xi,\pi/2)>0$ for all $\xi\geq\xi_0$  from a region S3~[S3s] where $\sigma_2 (\xi,\pi/2)\leq0$ in a finite interval $\xi_0<\xi_{\text{min}}\leq\xi\leq\xi_{\text{max}}$. The transition is given by looking at a local minimum $\xi^{*} \ge \xi_{0}$ such that
\begin{eqnarray}
&&\frac{d \sigma_2 }{d \xi} (\xi^*,\pi/2)=0,\label{eq:min_1}\\
&&\sigma_{2} (\xi^*,\pi/2)= 0.\label{eq:min_2}
\end{eqnarray}
Using Eq.~(\ref{Principal stress_y}), the condition~(\ref{eq:min_1}) gives
\begin{equation}
\xi^{*}= \cosh^{-1} \sqrt{\frac{3}{1+\delta^{2}}}. \label{Min sigma2}
\end{equation}
The solution given by Eq.~(\ref{Min sigma2}) should satisfy $\xi^{*} \ge \xi_0$. This condition holds for $\cosh\xi_0 \le  \sqrt{2}$, which is equivalent to require that $\delta \le \delta^{*} = 1/\sqrt{2}$.
The curve $\tau_{2}(\delta)$ is found by imposing condition~(\ref{eq:min_2}). Using Eqs.~(\ref{Principal stress_y},\ref{Min sigma2}) one finds
\begin{equation}
\tau_{2}(\delta) = 
\frac{2\sqrt{3}(2-\delta^{2})^{3/2} }
{2\sqrt{3}(2-\delta^{2})^{3/2}  - 9(1-\delta^{2})}, \text{ with } 0\leq\delta \le \frac{1}{\sqrt{2}}.
\end{equation}
One can show that the curves $\tau_{2}(\delta)$ and $\tau_{1} (\delta)$ intersect at $\left( \delta^{*}=1/\sqrt{2},\tau^{*}=2+\sqrt{2} \right)$. Hence, $\left( \delta^{*},\tau^{*} \right)$ corresponds to a triple point in the phase diagram linking the states S1, S2 and S3 (see Fig.~\ref{Fig: Diagram}). Furthermore, notice that he slopes of $\tau_1(\delta)$ and $\tau_2(\delta)$ coincide at the triple point ($\tau'_{2}(\delta^{*})=\tau'_{1} (\delta^{*})$). 

The critical curve $\tau_{s}(\delta)$ defines the transition from the states S2 and S3 to the states with slack regions S2s and S3s. As the slack regions concentrate close to the major axis of the elliptic hole (see Fig.~\ref{Fig: FiguresSheet}), it is sufficient to inspect the behaviour of $\sigma_1(\xi\geq \xi_0,0)$ as given by Eq.~(\ref{Principal stress_x}). Using the same methodology as for the determination of $\tau_2(\delta)$, one finds that the curve $\tau_{s}(\delta)$ is given by
\begin{equation}
\tau_{s} (\delta) = \frac{2 \sqrt{3}}{2 \sqrt{3} - 9 \delta (1-\delta^{2} )} , \text{ with } 0\leq\delta < \frac{1}{\sqrt{3}}.
\label{eq:taus}
\end{equation}
Eq.~(\ref{eq:taus}) shows that $\tau_s(\delta)$ diverges as $\delta \rightarrow 1/\sqrt{3}$ and $\tau_s(0)=\tau_{c}(0)=1$. Therefore, slack regions occur when the elliptic hole is slender. This latter result motivates the study of the stretched membrane in the crack-like limit.

As $\delta \rightarrow 0$ the elliptic hole becomes slender and tends to a straight cut of length $2a$, resembling a crack in a stretched membrane~\cite{mahmood2018cracks}. In this case the stress field close to the elliptic hole is better represented in polar coordinates ($r,\theta$) with origin $r=0$ at the crack tip $x=a$. We are interested in the asymptotic behaviour of the stress field in the vicinity of the crack tip. In the appendix, we show that the limits $r \rightarrow 0$ and $\delta \rightarrow 0$ do not commute. 

\begin{figure}[t]
\begin{center}
\includegraphics[width=\columnwidth]{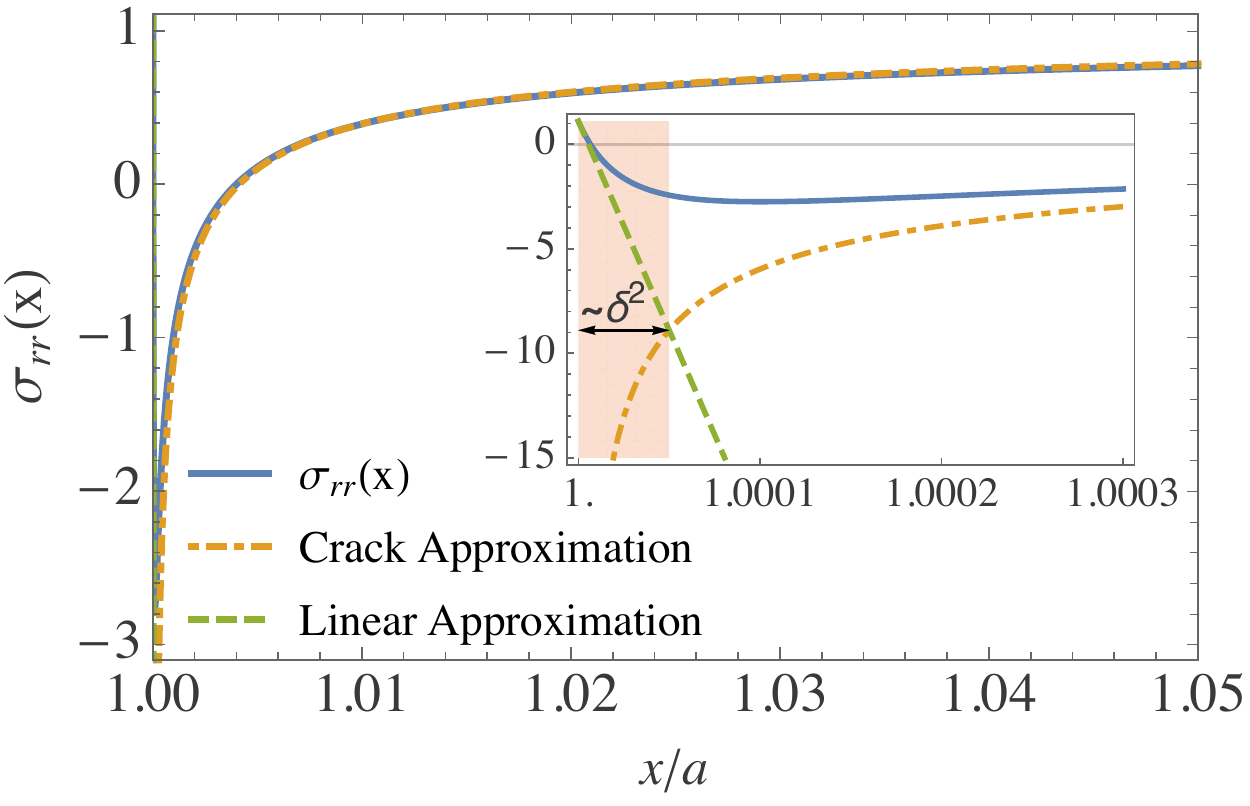}
\end{center}
\caption{\label{Fig: StressProfile}  Profile of $\sigma_{rr}$ along the $x$-axis close to the hole for $\delta=0.01$ and $\tau=1.1$. The linear and crack approximation given by Eqs.~(\ref{Linear approx},\ref{Crack approx}) are also shown. The inset is a zoom in the vicinity of the tip. The shaded area represents the region where the crack approximation fails.}
\end{figure}

\begin{figure*}[t]
\begin{center}
\includegraphics[width=\textwidth]{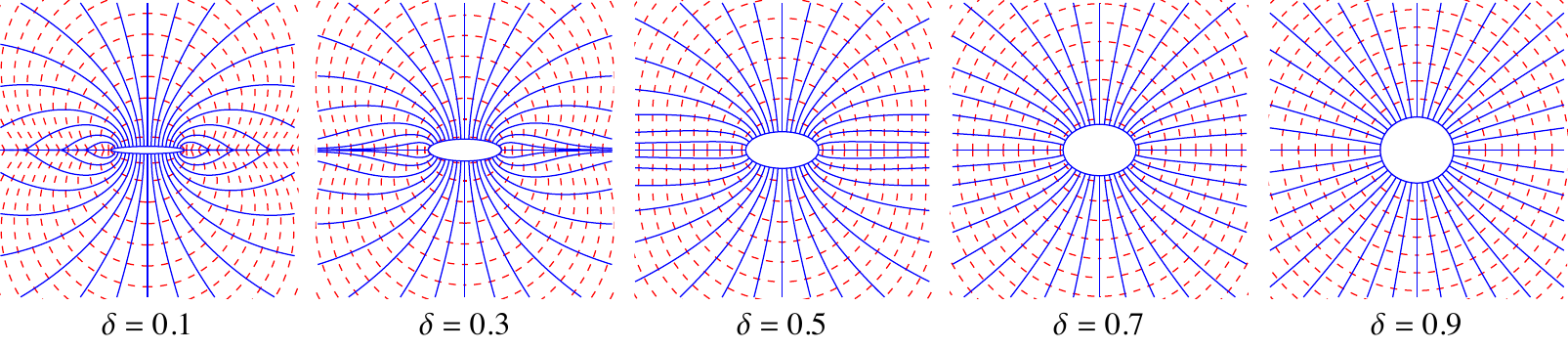}
\end{center}
\caption{\label{Fig: Principal directions}
The lines of tension defined by the local principal stress directions given by Eq.~(\ref{tan 2 beta}). Notice that the directions of the tension lines depend on the geometrical parameter $\delta$ only. Blue solid (resp. red dashed) lines correspond to the largest (resp. smallest) principal stress directions.
}
\end{figure*}

Let us first consider the asymptotic expansion of the stress field in powers of $r/a$ for a given nonzero value of $\delta$. As one approaches the crack tip ($r\rightarrow 0$) the radial stress component admits an expansion of the form (see the appendix)
\begin{equation}
\sigma_{rr} (r,0)=\tau T_{\text{out}} - \frac{2\Delta T}{\delta^{3}} \frac{r}{a}+\mathcal{O}\left( (r/a)^{2} \right). \label{Linear approx} 
\end{equation}
Notice that $\sigma_{rr}(0,0) = \tau T_{\text{out}}$, as dictated by the boundary conditions. On the contrary, when one takes the crack limit $\delta \rightarrow 0$ before performing the asymptotic expansion, the stress tensor admits an expansion of the form
\begin{equation}
\sigma_{rr}\big|_{\delta=0}  (r,0)= \tau T_{\text{out}} - \frac{\Delta T}{\sqrt{2}} \sqrt{\frac{a}{r}} + \mathcal{O} \left( \sqrt{r/a}\right). \label{Crack approx}
\end{equation}
Excepting  the term $\tau T_{\text{out}}$, Eq.~(\ref{Crack approx}) corresponds to the asymptotic expansion of the radial stress component of a stationary crack~\cite{williams1961bending}. Notice that this expansion violates the original boundary conditions~(\ref{Boundary condition 2}) at the hole. For small $\delta \neq 0$, the crack limit approximation looses validity as one moves closer to the tip (see~Fig. \ref{Fig: StressProfile}). The crack approximation breaks down inside a cohesive zone-like region of typical size $r_{c}$. The radius $r_c$ can be estimated either as the radial distance at the intersection of the linear approximation~(\ref{Linear approx}) with the crack approximation~(\ref{Crack approx}) or as the radial distance of the local minimum of $\sigma_{1} (r,0)$. The two criteria yield $r_{c}\sim a\delta^2$ which coincides, up to proportionality factors, both with the radius of curvature at the tip of the elliptic hole and with the radius of convergence of the series expansions of the stress field given by Eqs.~(\ref{Stress polar coordinates}) in the appendix.

\section{Discussion}

Fig.~\ref{Fig: Diagram} shows the phase diagram of the different accessible global stress configurations of a stretched sheet prior to any out-of-plane deformation. To release compressive stresses, a sheet with zero flexural rigidity buckles as soon as $\tau>\tau_c(\delta)$, implying that the states S3 and S3s of the phase diagram would not be accessible. However, a sheet with finite flexural rigidity should overcome a critical compressive stress to buckle and thus, one expects the whole phase diagram to be relevant. The present study shows that a slight deviation from the original Lam\'e problem (a change of the geometry of the perforated hole) gives rise to a phase diagram with different global stress states that might induce a rich variety of wrinkled patterns. We believe our model problem worths experimental studies to answer fundamental questions such as: Do wrinkled regions exhibit topologies similar to those of UT regions shown in Fig.~\ref{Fig: FiguresSheet}? To what extent the phase diagram computed in Fig.~\ref{Fig: Diagram} remains relevant as $\tau$ is increased?

From a theoretical point of view, the difficulties of predicting the state of the sheet beyond the pre-buckled one are due to three main observations: 
\begin{itemize}[label=$\bullet$,leftmargin=*]
\item UT regions exhibit complex, possibly disconnected, contours.  Indeed, Fig.~\ref{Fig: Diagram} shows that the Lam\'e case $\delta=1$ is peculiar as a small eccentricity modifies the nature of the transition to the wrinkled state. While for $\delta =1$ a single symmetrical wrinkled state S1 exists, a new intermediate state S2 (or S2s) emerges for $\delta \neq1$.
\item The eccentricity of the hole may generate slack regions whose effect on the post-wrinkling process is not documented neither experimentally nor theoretically.
\item The tension lines are not straight curving considerably as the hole becomes slender (see Fig.~\ref{Fig: Principal directions}). This feature hinders any prediction about the shape of wrinkles from the pre-buckled state.
\end{itemize}

These ascertainments prevent from performing classical NT or FFT analysis. Due to the absence of axial symmetry, it is not obvious how to perform a perturbation analysis around the planar state or to find the regions of the sheet where only tensile forces take place. We believe that an accurate description of a general wrinkling problem should be considered as a step-by-step dynamical-like problem, in the sense that once the membrane starts to buckle, the stress landscape on the whole membrane is modified and the buckled zones on the membrane are reshaped accordingly. In our specific problem, for $\delta <1$ and $\tau\gtrsim\tau_{c} (\delta)$, compressed regions occur mainly at the tips of the ellipse, consequently, the buckling instability will be preferably located around the tips of the ellipse, much as a crack-like problem. 

Specifically, wrinkles as wavy periodic structures could be peculiar patterns observed exclusively in symmetric configurations while the generic buckling instability would  induce in the first place folded patterns---localised out-of-plane excursions of the sheet separated by flat regions under pure tension~\cite{King9716}. This assumption opens alternative approaches for the description of the post-buckling behaviour. For example, if one considers the folding mechanism as a mean to suppress both normal and shear tractions along folds, one can envision the resulting pattern as traction-free crack lines. In this sense, one expects the physics underlying the selection of the folds pattern to be analogous to that of cracks and  then must be treated as a dynamical process. The extension of a single fold should satisfy the principle of local symmetry~\cite{gol1974brittle} and a Griffith energy criterion~\cite{broberg1999cracks} with vanishing fracture energy. Thus, the ``equations of motion''  of the fold can be written as $K_{II}=0$ and $K_I=0$, where $K_I$ (resp. $K_{II}$) is the mode~I (resp. mode~II) stress intensity factor associated to the square root singularity of the stress field at the tip of the fold. These conditions ensure that the stress field in the periphery of the fold is shear-free and tensionless. The proposed equations  allow for predicting both the shape and the extension of a single fold. In a realistic situation with many folds, one should supplement the equations of motion of each fold with a global elastic energy functional of the sheet, whose minimisation selects the geometrical and topological properties of the folding pattern.

Finally, we have studied the stress field in a sheet with an elliptic hole subjected to a differential tension between its inner and outer boundaries. We have found that regions around the hole can be either in a taut, slack or UT state. This yields a rich phase diagram of different global stress states that should have an effect on the three-dimensional shape of the sheet beyond the buckling instability. These results demonstrate that slight geometrical asymmetries or inhomogeneous loading conditions might lead to complex wrinkled patterns. Our study calls for experiments to probe the robustness of the wrinkling paradigm.

\acknowledgments
I. A.-S. acknowledges the financial support of CONICYT DOCTORADO BECAS CHILE 2016-72170417.

\begin{widetext}

\renewcommand{\theequation}{A-\arabic{equation}}
\setcounter{equation}{0}
\section{Appendix: Asymptotic expansion of the stress field}

The components of the stress field in polar coordinates ($r,\theta$) can be obtained from Eqs.~(\ref{stresses}) by using the following transformations
\begin{subequations}
\begin{eqnarray}
\sigma_{r r}  + \sigma_{\theta \theta} &=& \sigma_{\xi\xi} + \sigma_{\eta\eta}, \\
\sigma_{rr} -\sigma_{\theta \theta} - 2i\sigma_{r \theta} &=& e^{2i(\theta-\alpha)} (\sigma_{\xi\xi} -\sigma_{\xi\eta} -  2i\sigma_{\xi\eta}).
\end{eqnarray}
\end{subequations}
where the angle $\alpha$ is defined by $e^{2i\alpha} = \sinh(\xi+i\eta) / \sinh(\xi-i\eta)$. Let us define the tip $x=a$ of the elliptic hole as the origin $r=0$.
Using Eq.~(\ref{eq:transf}) and the stress field given by Eqs.~(\ref{stresses}), one obtains 
\begin{subequations}
\begin{eqnarray}
\sigma_{rr} + \sigma_{\theta \theta} &=& 2 \tau T_{\text{out}} - 2 \Delta T\, \text{Re} \left[ \frac{a+r e^{i\theta}}{\sqrt{r^{2} e^{2i\theta} +2r a e^{i\theta} +a^{2}\delta^{2}}} \right] , \label{Stress polar coordinates 1} \\
\sigma_{rr}-\sigma_{\theta \theta}-2i\sigma_{r \theta} &=& \Delta T\, \frac{2 a^{3}\delta^{2} e^{2i\theta}- \left( (1-\delta^{2}) e^{i \theta} - (1+\delta^{2}) e^{3 i \theta}\right) a^{3} r}{\left( r^{2} e^{2i\theta} +2r a e^{i\theta} +a^{2}\delta^{2}\right)^{3/2}}.  \label{Stress polar coordinates 2} 
\end{eqnarray} \label{Stress polar coordinates}
\end{subequations}
The asymptotic expansion of the stress field in the vicinity of the tip of the slender ellipse can be performed in two ways: either by assuming a finite $\delta$ and taking $r\rightarrow 0$ or by taking first $\delta=0$ in  and then $r\rightarrow 0$. Using Eqs. (\ref{Stress polar coordinates}), one gets for the former case
\begin{subequations}
\begin{eqnarray}
 \sigma_{rr} &=& \tau T_{\text{out}} - \frac{\Delta T }{\delta} \left(1 - \cos 2\theta \right)
+ \frac{\Delta T}{2 \delta^{3}}  \frac{r}{a}\left(  \left( 1 - \delta^{2} \right) \cos \theta - \left( 5 - \delta^{2}  \right) \cos 3\theta \right) + \mathcal{O}\left( (r/a)^{2} \right), \\
\sigma_{\theta \theta}&=& \tau T_{\text{out}} - \frac{\Delta T }{\delta} \left(1 + \cos 2\theta \right)
+ \frac{\Delta T}{2 \delta^{3}}  \frac{r}{a} \left( 3 \left( 1 - \delta^{2} \right) \cos \theta + \left( 5 - \delta^{2}  \right) \cos 3\theta \right) +  \mathcal{O}\left( (r/a)^{2} \right), \\
\sigma_{r \theta} &=& - \frac{2 \Delta T}{\delta} \cos{\theta} \sin{\theta} 
+\frac{\Delta T}{\delta^{3}}  \frac{r}{a} \left( 3 - \delta^{2} + \left(5 - \delta^{2} \right) \cos 2\theta \right) \sin \theta +  \mathcal{O}\left( (r/a)^{2} \right).
\end{eqnarray}
\end{subequations}
In the crack limit case ($\delta=0$), the asymptotic expansion of Eqs.~(\ref{Stress polar coordinates}) read
\begin{subequations}
\begin{eqnarray}
 \sigma_{rr} \big|_{\delta=0} &= & \tau T_{\text{out}}
 - \frac{ \Delta T }{2\sqrt{2}} \sqrt{\frac{a}{r}} \cos{\frac{\theta}{2}} \left( 3 -\cos{\theta}\right) 
 + \mathcal{O} \left( \sqrt{\frac{r}{a}}\right), \\
\sigma_{\theta \theta} \big|_{\delta=0} &=& \tau T_{\text{out}} 
- \frac{ \Delta T }{\sqrt{2}} \sqrt{\frac{a}{r}} \cos^{3}{\frac{\theta}{2}}
+ \mathcal{O} \left( \sqrt{\frac{r}{a}}\right), \\
\sigma_{r \theta} \big|_{\delta=0} &=& - \frac{\Delta T}{2 \sqrt{2}} \sqrt{\frac{a}{r}}\cos{\frac{\theta}{2}} \sin{\theta}
+ \mathcal{O} \left( \sqrt{\frac{r}{a}}\right).
\end{eqnarray}
 \end{subequations}
 
 \end{widetext}
 
\bibliographystyle{eplbib}
\bibliography{Article}

\end{document}